\documentclass[12pt]{iopart}

\newcommand{\pp}{{$p-p$~}}

\usepackage{graphicx}
\usepackage{color}
\begin{document}

\title{Prospects for Strangeness Production in \pp Collisions at LHC}

\author{I.~Kraus$^{1}$, J.~Cleymans$^{1,2}$, H.~Oeschler$^{1}$ and K.~Redlich$^{3,4,5}$}
\address{$^{1}$Institut f\"ur Kernphysik, Darmstadt University of Technology, D-64289 Darmstadt, Germany}
\address{$^{2}$UCT-CERN Research Centre and Department  of  Physics, University of Cape Town, Rondebosch 7701, South Africa}
\address{$^{3}$Institute of Theoretical Physics, University of Wroc\l aw, Pl-45204 Wroc\l aw, Poland}
\address{$^{4}$GSI Hemholtzzentrum f\"ur Schwerionenforschung, D-64291 Darmstadt, Germany}
\address{$^{5}$Fakult\"at f\"ur Physik, Universit\"at Bielefeld, D-33501 Bielefeld, Germany}

\ead{Ingrid.Kraus@cern.ch}
\begin{abstract}

Prospects for strangeness production in \pp collisions at the Large
Hadron Collider (LHC) are discussed within the statistical model.
Firstly, the system size and the energy dependence of the model
parameters are extracted from existing data and extrapolated to
LHC energy. Particular attention is paid to demonstrate that the
chemical decoupling temperature is independent of the system size.
In the energy regime investigated so far, strangeness production
in \pp interactions is strongly influenced by the canonical
suppression effects. At LHC energies, this influence might be
reduced.
Particle ratios with particular sensitivity to canonical effects
are indicated.

Secondly, the relation between the strangeness production and the
charged-particle multiplicity in \pp interactions is
investigated. In this context  the multiplicity dependence studied
at Tevatron is of particular interest.
There, the trend in relative strangeness production known from centrality dependent heavy-ion collisions is not seen in multiplicity selected \pp interactions.
However, the conclusion from the Tevatron measurements is based on
rather limited  data samples with low statistics and number of
observables. We argue, that there is an absolute need at LHC to
measure strangeness production in events with different
multiplicities to possibly disentangle relations and differences
between particle production in \pp and heavy-ion collisions.

\end{abstract}

\pacs{12.40.Ee, 25.75.Dw}
\vspace{2pc}
\noindent{\it Keywords}: Statistical model, Strangeness undersaturation, Particle production
\maketitle

\section{Introduction} %----------------------------------------------------------------------
For more than half a century the statistical model has been used
to describe particle production in high-energy
collisions~\cite{fermiheisenberghagedorn}. In the past it has
evolved into a very useful and successful model describing a large
variety of data.  In particular, hadron yields in central
heavy-ion collisions have been described in a very systematic and
appealing way unmatched by any other model~\cite{pbpb}. The
statistical model has also provided a very useful framework for
the centrality and system-size dependence of particle
production~\cite{syssize,bec}. The applicability  of the model in
small systems like \pp and $e^+-e^-$ annihilation has been the
subject of several recent publications~\cite{pppred,ee2}.
In the following, we present  predictions of  the statistical
model for particle productions  at the LHC energy. Any deviation
of data from our predictions allows to disentangle a  new physics
phenomena  in high energy \pp collisions.

The thermal parameters of central heavy-ion collisions, given by
the temperature $T$ and the baryon chemical potential $\mu_B$,
appear to fall on a common chemical freeze-out curve in the
T--$\mu_B$ plane.
This observation  allows to
extrapolate the model parameters to the LHC energy regime and to
predict particle ratios in heavy-ion collisions. To verify the
validity of this method for \pp interactions we investigate the
system-size dependence of the model parameters. In
Section~\ref{sec-sys} we focus on differences emerging   in the
chemical freeze-out temperatures in Refs.~\cite{syssize} and
\cite{bec} as recently studied in Ref.~\cite{new}.

The canonical suppression of the strange particle phase-space in
\pp collisions was found  to be essential to quantify the  SPS and
RHIC data. Its evolution to LHC energy is necessary to correctly
describe the strangeness sector. The energy dependence of the
canonical effect is discussed in detail in Section~\ref{sec-extra}.
Here, we propose to distinguish
between early and late strangeness production from experimental
data based on the strength of the canonical suppression.

As a first step towards higher collision energies, we consider
Tevatron data in Section~\ref{sec-tev}. We argue that, the
strangeness production and in particular its multiplicity
dependence might suggest that the production is established at an
early stage. However, as we discuss in the Outlook, this
observation  is based on a rather weak ground and only experiments
at LHC energies might  provide   definite conclusions.

\section{\label{sec-sys}System size and centrality dependence} %-----------------------------------------------
\begin{figure}
\begin{minipage}[b]{0.6\linewidth}
\includegraphics[width=0.99\linewidth]{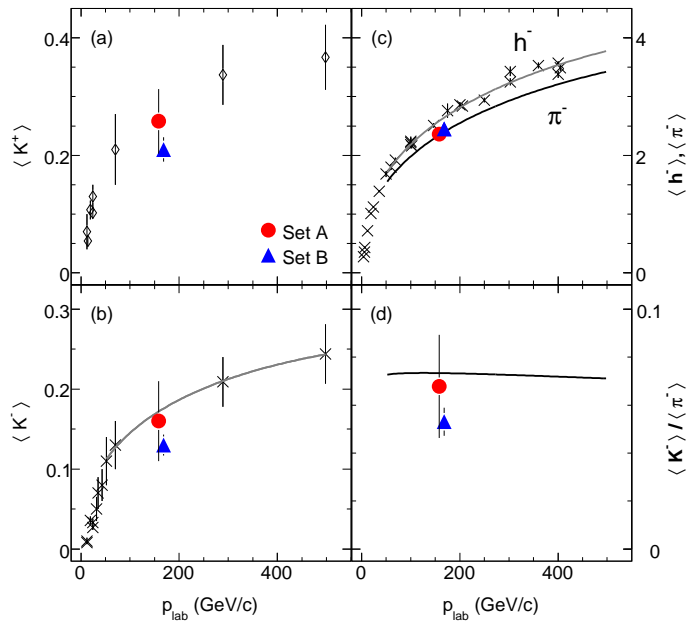}
\end{minipage}\hfill
\begin{minipage}[b]{0.4\linewidth}
\caption{\label{fig1} Charged kaon yields (left panels),
negatively charged hadron $h^-$ and pions $\pi^-$ (c), and
$K^-/\pi^-$ ratios (d) in \pp collisions as a function of
laboratory momentum. The charged kaon yields, $K^+$ (diamonds) and $K^-$
(crosses), are from Ref.~\cite{compK}. The lines are fits
to data. The SPS yields from Ref.~\cite{dataSPSK} (circle)  and
from Ref.~\cite{dataSPSK0s} (triangle) are also shown. The
negatively charged hadrons are  from  Ref.~\cite{comph}. }
\end{minipage}
\end{figure}
The trends of temperature and baryon chemical potential when
varying the number of participating nucleons from central heavy
ion to \pp collisions were studied at SPS~\cite{syssize,bec} and
RHIC~\cite{rhic}. The same value of $\mu_B$ is found in
phase-space integrated data, while mid-rapidity data show a
decreasing number of net baryons as the multiplicity decreases.
This, together with smaller $\mu_B$ at higher energy, can be
explained by weaker stopping due to higher transparency.
On the other hand, the chemical freeze-out temperature was found
to be multiplicity-independent at RHIC and also in our   analysis
at SPS ~\cite{syssize}. These results are in contrast with
findings from Ref.~\cite{bec} where the temperature was shown to
be higher in smaller systems. Such behavior has dramatic
consequences, e.g., it indicates  that  one can probe in \pp
interactions QCD matter beyond the freeze-out line
established in heavy-ion collisions~\cite{na61}. Below we discuss what is  the
origin of such different  predictions.

\begin{figure}
\begin{minipage}[b]{0.5\linewidth}
\includegraphics[width=\linewidth]{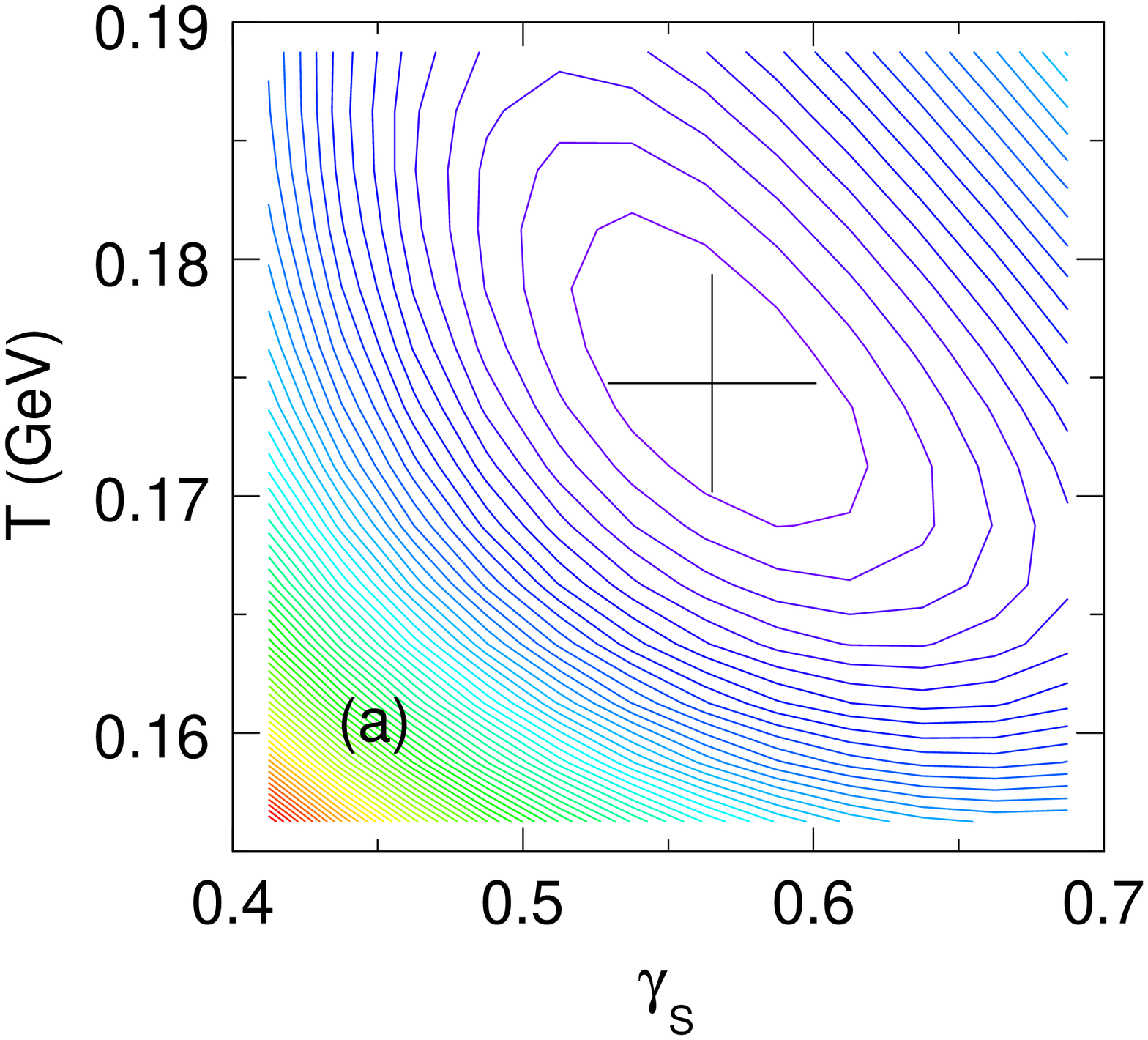}
\end{minipage}
\begin{minipage}[b]{0.5\linewidth}
\includegraphics[width=\linewidth]{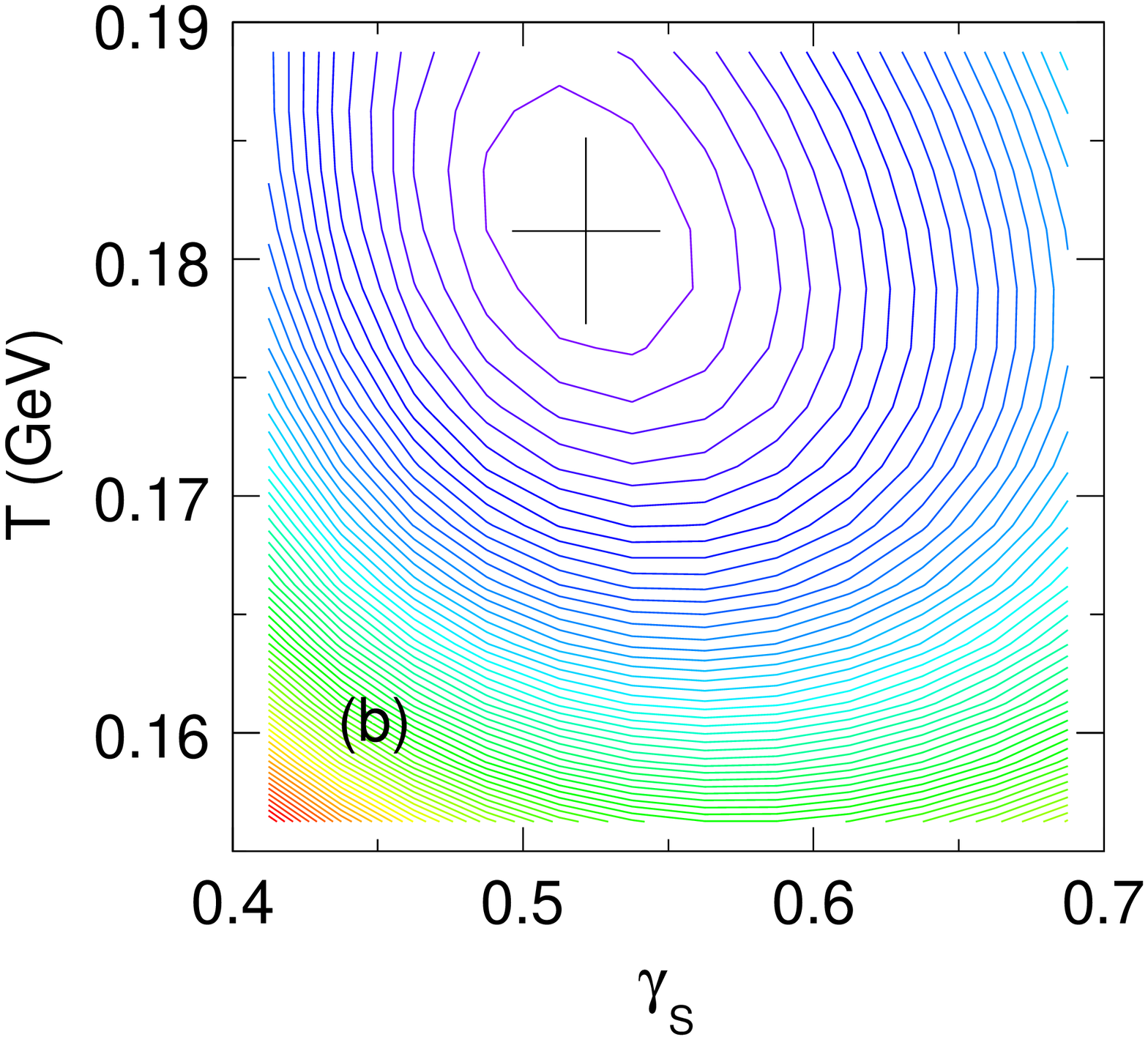}
\end{minipage}
\caption{\label{fig2} The $\chi^2$ scan  in the
(T--$\gamma_S$)-plane. Starting from its  minimum, $\chi^2$
increases by 2 for each contour line. Fit to (a) data set of
Ref.~\cite{pppred} (data set {\sl A1} in \cite{new}) and (b) same
data set but charged kaons as used in Ref.~\cite{bec} (data set
{\sl A4} in \cite{new}), in the canonical model. The minima are
indicated by the crosses.}
\end{figure}

\begin{figure}
\begin{minipage}[b]{0.5\linewidth}
\includegraphics[width=\linewidth]{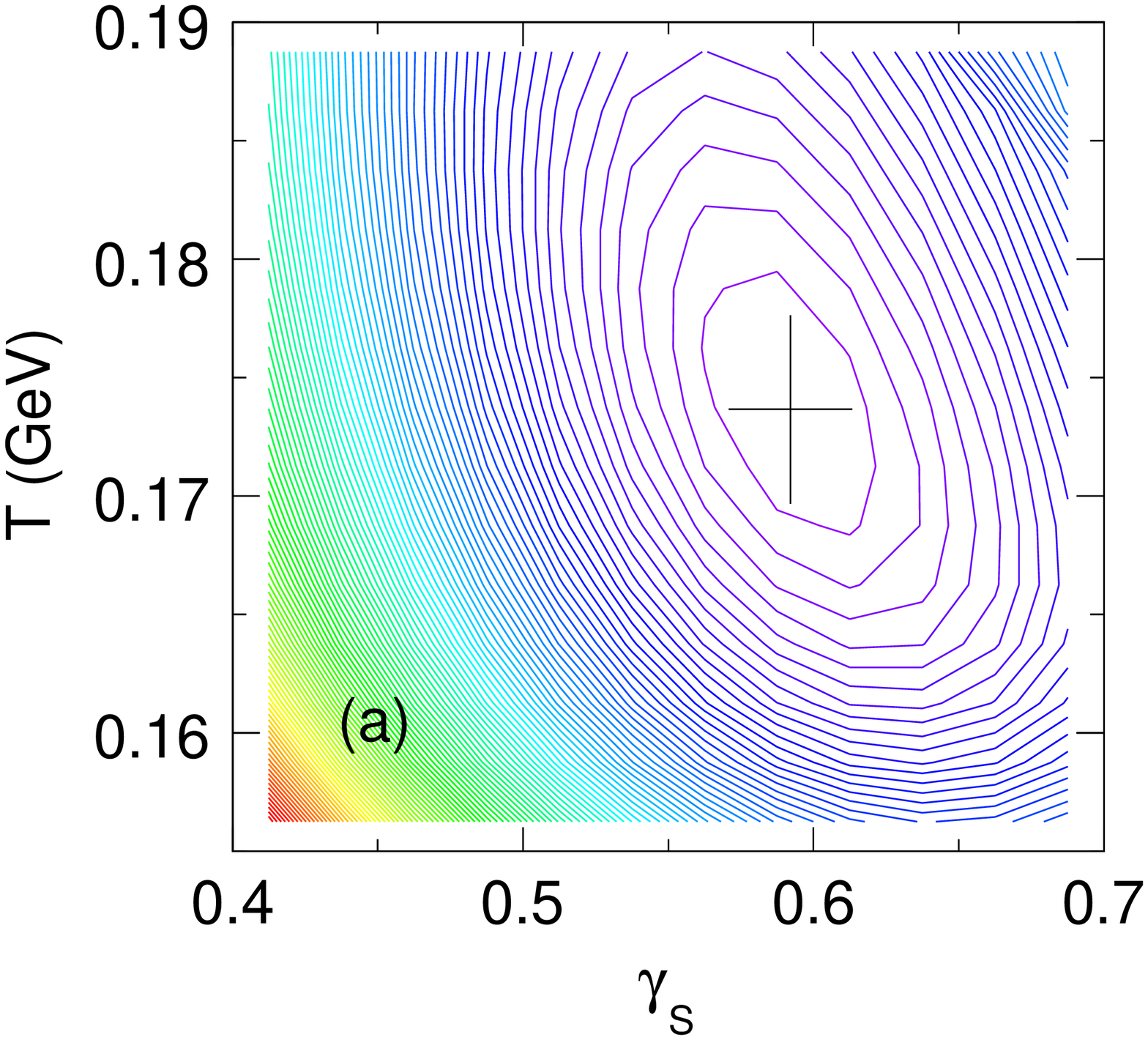}
\end{minipage}
\begin{minipage}[b]{0.5\linewidth}
\includegraphics[width=\linewidth]{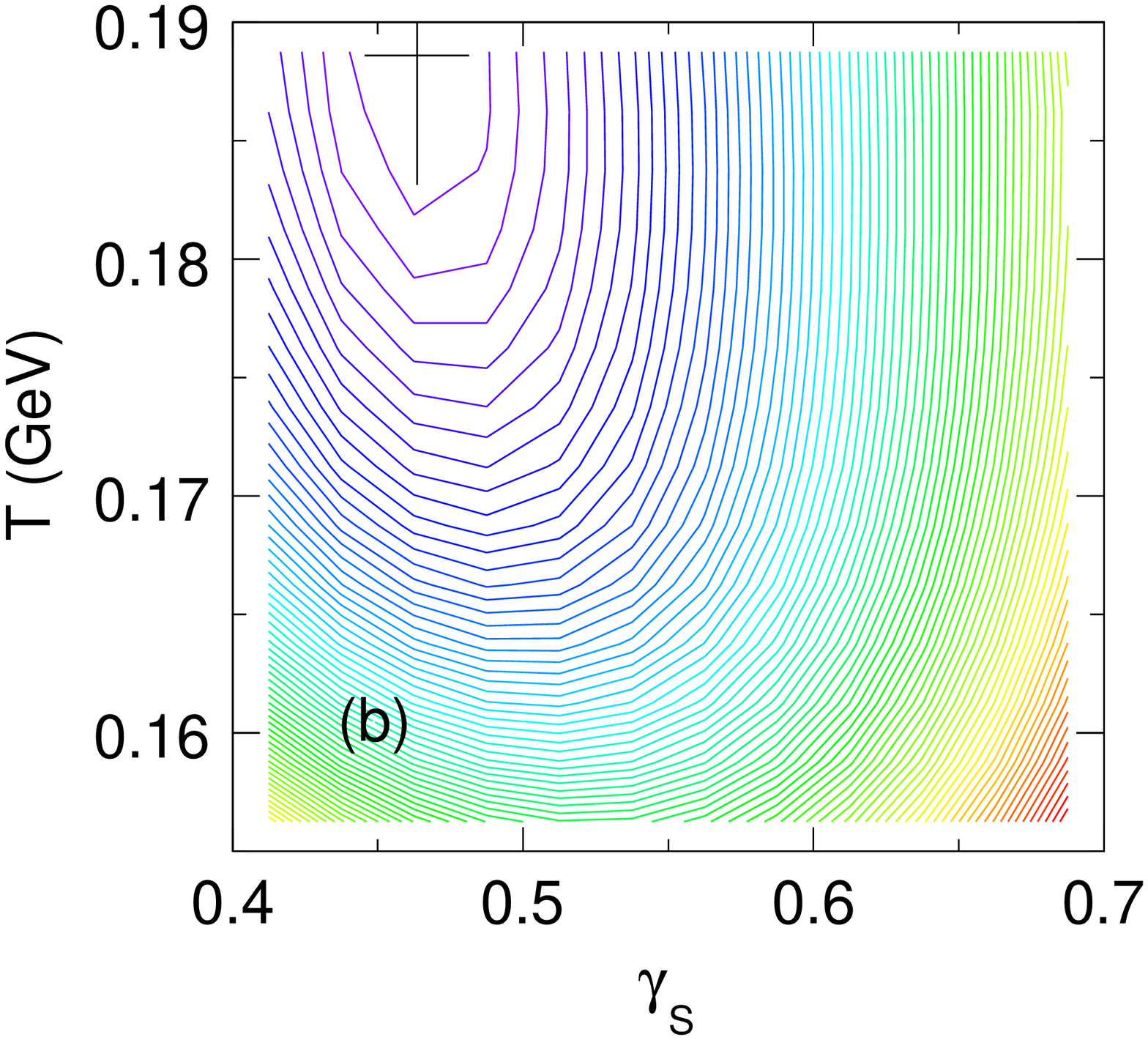}
\end{minipage}
\caption{\label{fig3} Same data set as in Fig. \ref{fig2}(a) but
(a) extended by $\Xi$ and $\Omega$ baryons and (b) by the $\phi$
meson, respectively.}
\end{figure}

The largest difference in extracted model parameters from Refs.~\cite{syssize}
and \cite{bec} appears in \pp collisions. The main reason is
that the above  analyses are based on different data sets and in
some cases also on  different numerical values for particle
yields. The largest differences  are observed in the charged kaon
yields as  illustrated in  Fig.~\ref{fig1}. Some data appear below
the line which has been obtained from an interpolation between
different lab momenta. The use of those data in the statistical model fit imply a stronger
suppression of the strange-particle phase-space (smaller
$\gamma_S$). This has to be compensated by higher $T$ in order to
describe various strange-particle yields like  e.g.~hyperons and
K$^0_S$ mesons. This is illustrated in Fig.~\ref{fig2} where the
left panel shows the fit to the data set of Ref.~\cite{pppred} while in the
right panel the charged kaon yields  are replaced by values used
in Ref.~\cite{bec}. The charged kaons  in Fig.~\ref{fig2}(a) are
taken from the parametrisation proposed in Ref.~\cite{dataSPSK}.
Figure~\ref{fig3}(a)  shows that these data are consistent with
preliminary data obtained for yields of multi-strange baryons.

The analysis in Ref.~\cite{bec} includes the $\phi$ meson. Based
on measured data we have shown in Ref.~\cite{syssize} that the
$\phi$ meson behaves like particle with strangeness between 1 and
2 in its multiplicity dependence. Therefore, this particle  cannot be correctly treated  in  the
statistical-model analysis. When including the $\phi$ meson in the analysis, we
are unable to  find a set of parameters that describes all
particle yields  simultaneously.  In addition, the fit results in
large $\chi^2$ along with the values of thermal parameters that
leave a reasonable range as  demonstrated in Fig.~\ref{fig3}(b).
Only fits that contain the $\phi$ meson result in a very high
temperature that exceeds that obtained in central heavy-ion
collisions at the corresponding energy.

From the above we conclude, that at a given energy the chemical
freeze-out temperature is system-size independent both at  RHIC
and at the SPS. In addition,  the  magnitude and the centrality
dependence of the   $\phi$ meson  is obviously inconsistent with
the  statistical model systematics. Consequently,  including this
particle in the fit can lead to misleading conclusions on the
model parameters and their qualitative  properties.

\section{\label{sec-extra}Extrapolation to LHC} %--------------------------------------------------------------
\begin{figure}
\begin{minipage}[b]{0.5\linewidth}
\includegraphics[width=\linewidth]{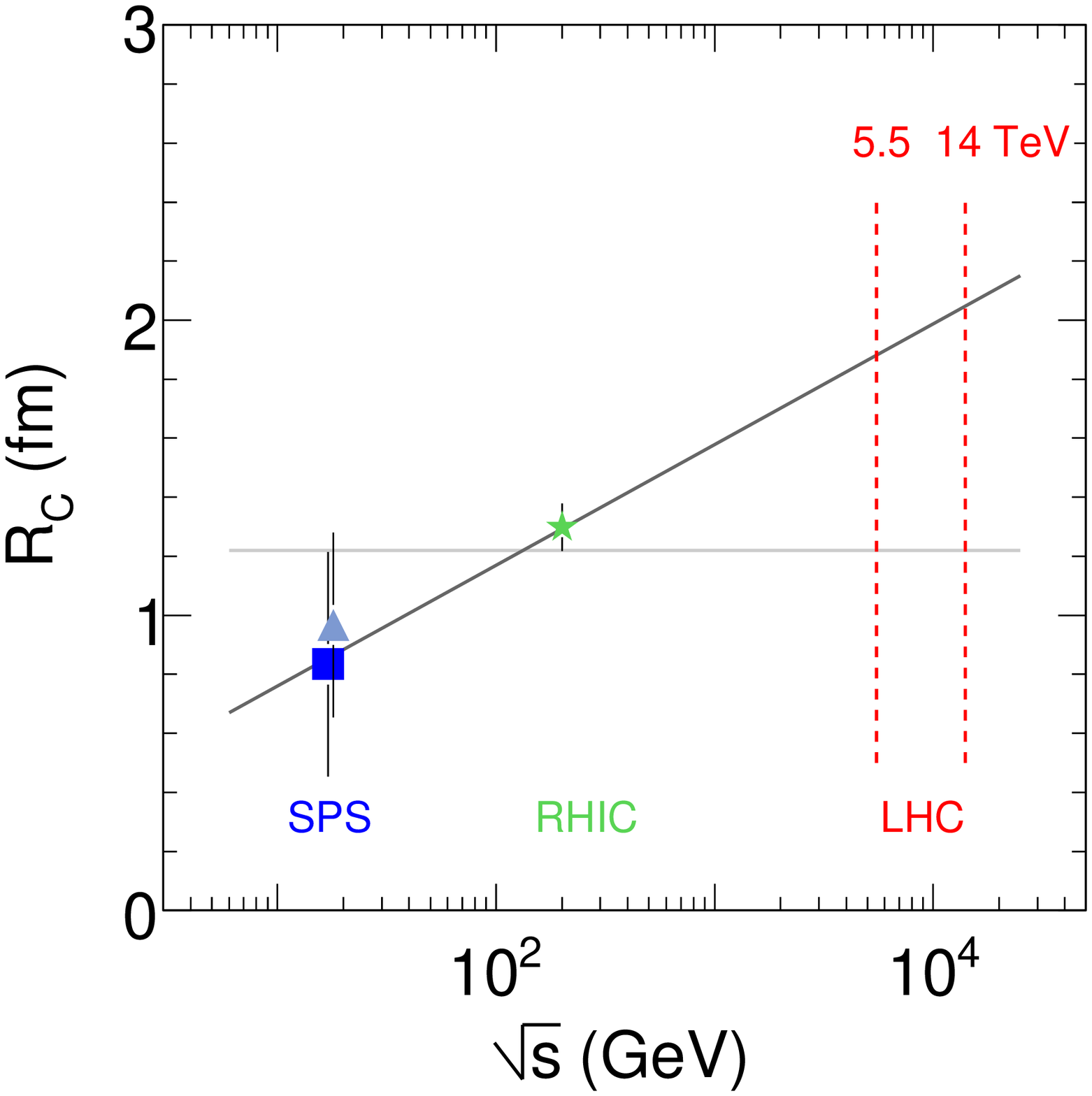}
\end{minipage}
\begin{minipage}[b]{0.5\linewidth}
\includegraphics[width=\linewidth]{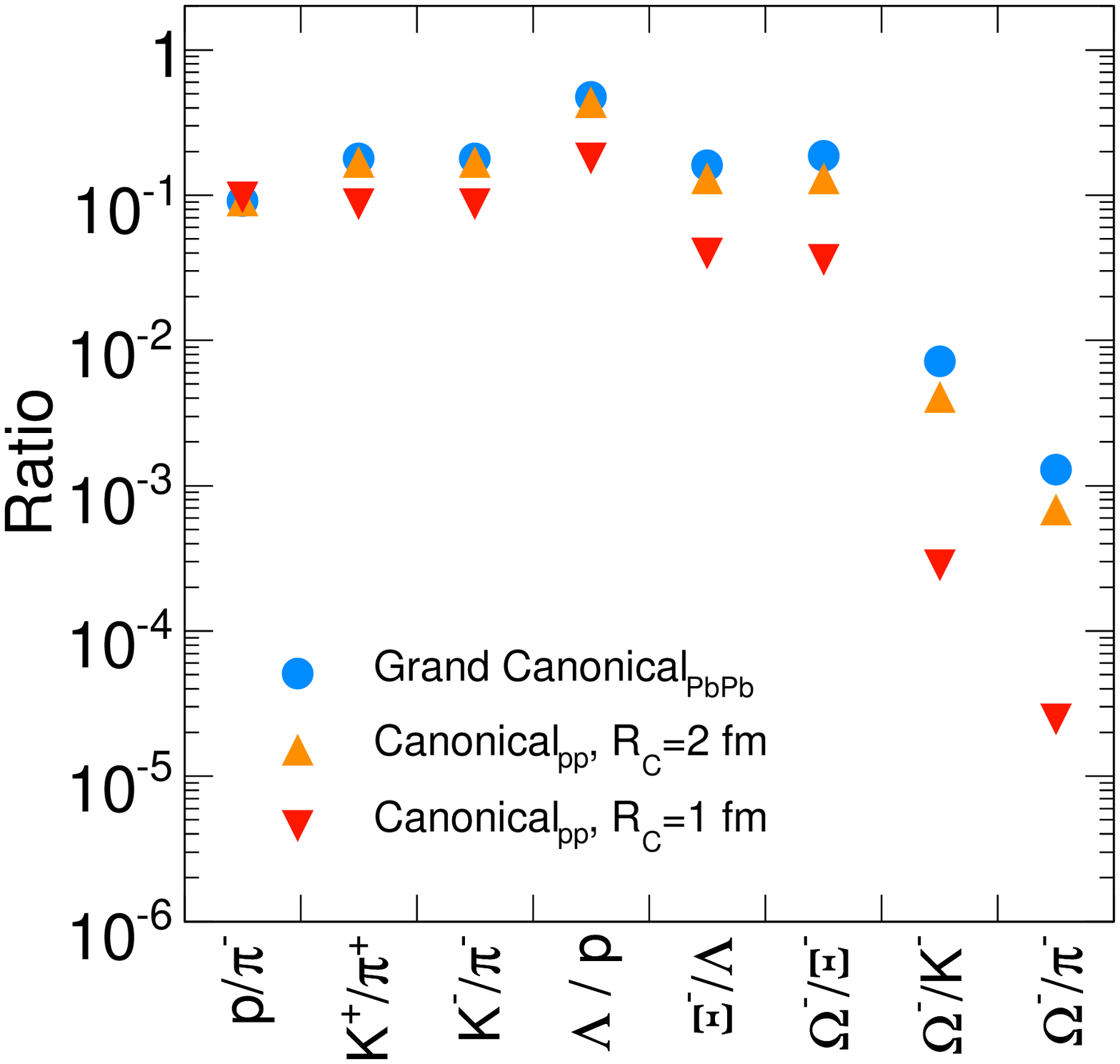}
\end{minipage}
\caption{\label{fig4} Left panel: The cluster radius $\rm R_C$ as
a function of energy, extracted from the analysis of SPS data
(triangle, square) and RHIC data (star)~\cite{pppred}. The lines
illustrate possible evolutions toward LHC energies as discussed in
the text. Right panel: Predictions for various particle ratios
using different values of $\rm R_C$~\cite{pppred}.}
\end{figure}
Based on the discussion above, we assume that the
multiplicity-independence of the temperature found at SPS and RHIC
is also valid at LHC.  In addition, to extrapolate  $T$ and
$\mu_B$ to  \pp interactions  at the LHC  we use the same
parametrisation of their  energy dependence  as it was found in
the systematics of heavy-ion collisions.   The baryon chemical
potential at midrapidity decreases towards smaller systems,
however, at the LHC $\mu_B$ is so  small that any variation
with system size can be neglected.

The canonical suppression is implemented in the following way.
Strangeness production is assumed to be correlated inside
chemically equilibrated clusters  that may be smaller than the
fireball. The size of these subvolumes (clusters) quantifies  the
strength of the canonical suppression. The correlation length
$R_C$ extracted from SPS and RHIC data, as illustrated in
Fig.~\ref{fig4} left, does not allow to quantify   its energy
dependence. We consider two extreme scenarios: (i) Small
subvolumes independent of the incident energy which can be
interpreted as being driven by the initial size of the proton or the
range of strong interactions. Thus strangeness production would
be determined at an early stage of the reaction.
(ii) Subvolumes that increase
with energy. In this case the strangeness production in \pp
collisions per charged particle would be multiplicity dependent,
reflecting a late determination of the strangeness content.
In such a scenario, at a given energy, one expects less strangeness
suppression in high multiplicity events. Consequently, high-multiplicity events
in \pp interactions could appear like heavy-ion collisions in
respect to the observed particle ratios.

As seen in Fig.~\ref{fig4} left, our actual knowledge from the SPS
and RHIC energies does not allow for  a definite  extrapolation of
the correlation volume towards the LHC energy. The radius $R_C$ of
the canonical correlation volume can appear  in the range between
1 and 2~fm.  Figure~\ref{fig4} (right) shows the sensitivity of
different particle  ratios to the value of the correlation radius.
From this figure it is clear, that measurements of   the $\Omega /
\pi$ and $\Omega / \rm K$ ratios are best  be used to extract the
strangeness correlation radius that quantifies canonical
suppression effects and the production dynamics of  strange
particles.
\section{\label{sec-tev} Strangeness production at Tevatron} %--------------------------------------------------------------
Before LHC results become available, Tevatron data provide
already some insight into strangeness production at higher
energies. There,  mainly kaons and $\Lambda$ hyperons were
measured as strangeness carriers. The kaon and pion yields were
found to increase smoothly and similarly with energy.  Thus, the
K$/\pi$ ratio seems to saturate with energy~\cite{tevatron}. This hints towards
saturating correlation length since the temperature is almost
unchanged  with collision energy beyond the SPS. When studying the
multiplicity dependence at  fixed energy, we can rule out a
possible counterbalance between T and $\rm R_C$. Indeed, in
$p-\bar{p}$ interactions the $p/\pi$  and K$/\pi$ ratios were
found to be independent of the charged-particle multiplicities.
These data indicate, that neither the  temperature nor the cluster
size exhibits variations  with multiplicity. This
interpretation is also supported by the multiplicity-dependent
$\Lambda / p$ ratio measured by the E735 and the CDF
collaboration.
\begin{figure}
\begin{minipage}[b]{0.5\linewidth}
\includegraphics[width=\linewidth]{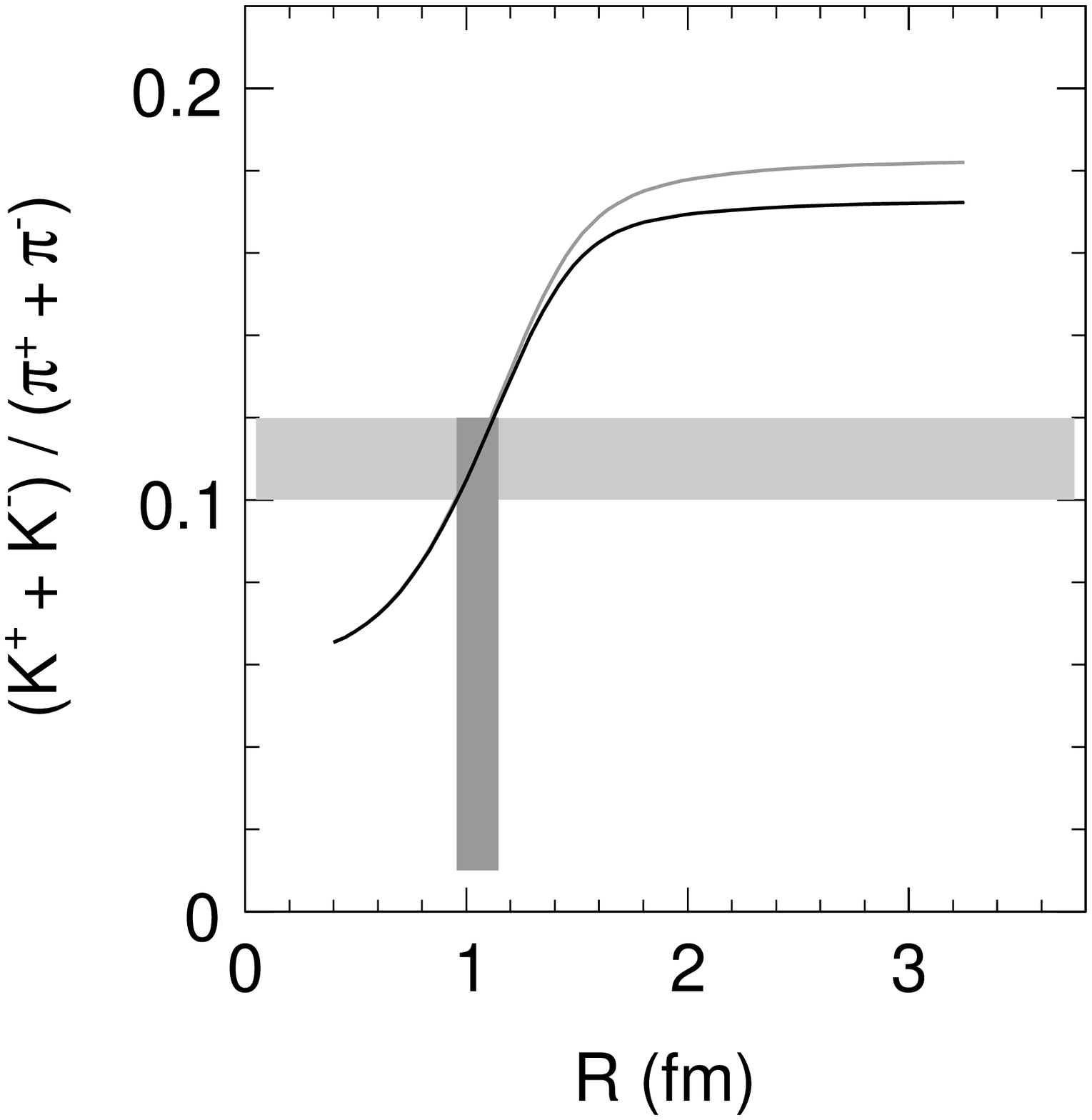}
\end{minipage}
\begin{minipage}[b]{0.5\linewidth}
\includegraphics[width=\linewidth]{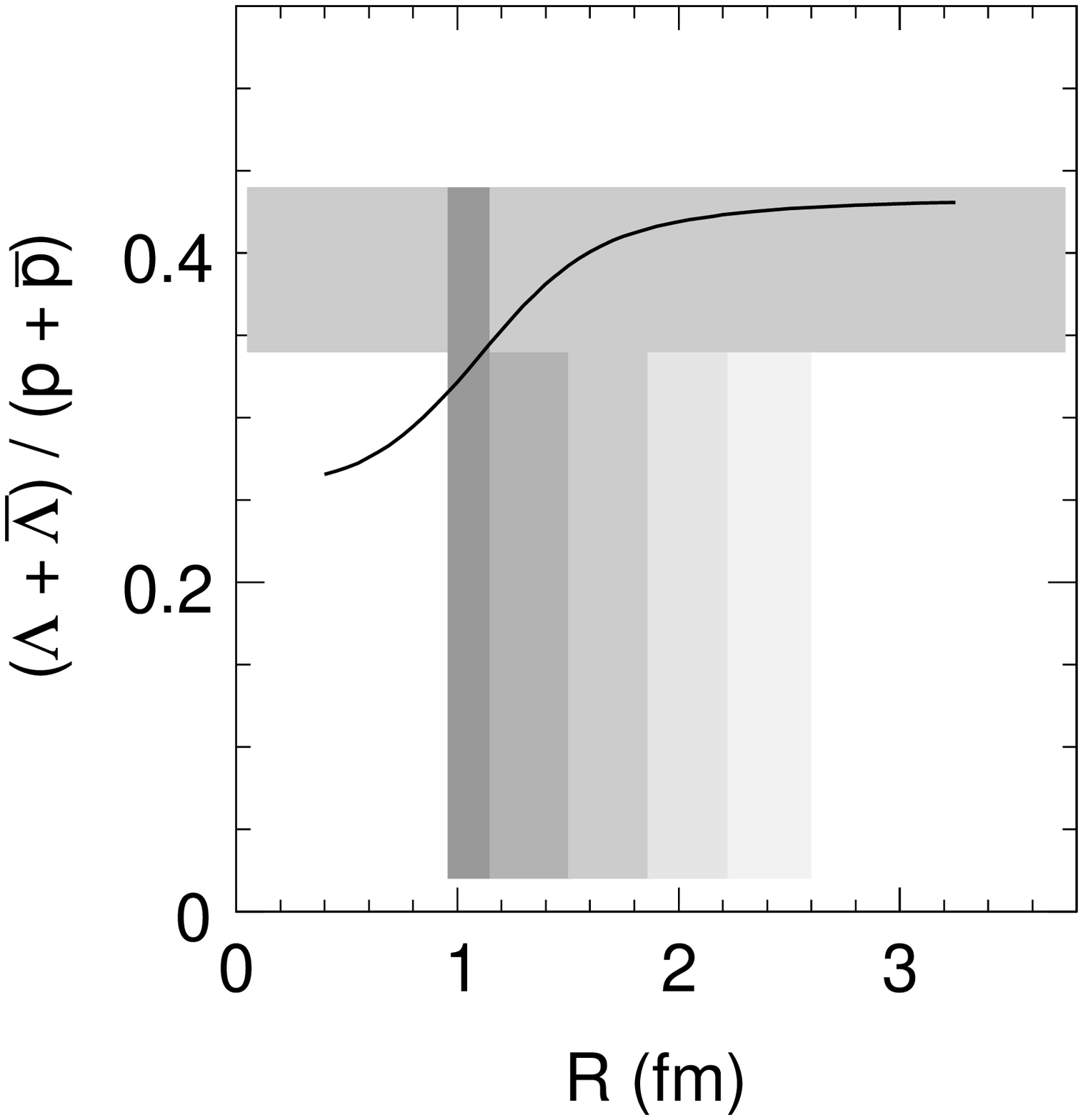}
\end{minipage}
\caption{\label{fig5} The statistical model results for  the
K/$\pi$ (left panel) and the $\Lambda/p$  (right panel) ratio
obtained with  the chemical freeze-out parameters expected at
Tevatron  ($T$ = 170 MeV and the initial quantum numbers  B = S =
Q = 0) as a function of the radius R of a  spherical volume.
Experimental results from Ref. \cite{tevatron} are indicated by
horizontal bands. Derived ranges of the correlation length are
displayed by vertical bands; for the $\Lambda/p$ ratio only a
lower limit can be extracted, the dark vertical band shows the range
extracted from the K/$\pi$ ratio. The gray curve in the left panel
indicates the uncertainties due to a mismatch between feed-down
corrections in data and model calculations.}
\end{figure}

After the above  qualitative discussion, we show in
Fig.~\ref{fig5}  a quantitative comparison of Tevatron data to the
statistical model. The measured K$/\pi$ ratio points to
significant canonical suppression. The cluster size is small with
none  or weak energy dependence  when comparing to RHIC and SPS
results. This supports the interpretation that  strangeness
production is determined  at an  early stage and is quantified by
the size of the colliding proton  or the range of strong
interactions. Consequently,  we expect strange/non-strange
particle ratios to saturate towards LHC energies with  only small
variations due to decreasing $\mu_B$. Due to the large errors the
$\Lambda /$p ratio contributes weakly to constrain the strength of the
canonical suppression.
\section{Summary and Outlook} %---------------------------------------------------
We have shown that the statistical model results are indeed very
sensitive to the data selection. We have traced back the apparent
variation of the chemical freeze-out temperature at SPS to kaon
yields that are not in line with earlier measurements at lower and
higher beam momenta. We conclude that the chemical freeze-out
temperature exhibits no system-size nor multiplicity dependence
in the energy range under study.
Thus, we confirm a similar observation made at RHIC.

The strangeness correlation length describes the strength of the
canonical suppression. Its energy and
multiplicity dependence can give insight into the question whether
the relative strangeness production is defined at an early or late
stage of the reaction. At present, the  extrapolation of the
correlation length extracted from SPS and RHIC data towards  LHC
energies is not unique due to rather large uncertainties. The
results from the Tevatron point towards small clusters  and an
early stage of strangeness production. However, this conclusion
is based on a rather incomplete and low-statistics data.

To quantify strangeness production and its energy dependence data
with  much better statistics are  needed to reduce errors.
Also there is a lack of multi-strange baryons, and the multiplicity
dependence of strangeness production at fixed  energy in \pp collisions
requires more complete data. Further, the E735 and CDF data cover a
charged particle densities up to 25 and 35, respectively. This
compares to very peripheral heavy-ion interactions, e.g.~it corresponds to
Cu-Cu collisions at RHIC where the 40$\%$ most central events were
left out.

At LHC we expect to have soon good statistics of events with a
charged-particle density up to 100 or more   in \pp interactions
corresponding  to semi-central Cu-Cu collisions. It is clear
that after a long period of successful running with \pp at the LHC
we will be able to  investigate the similarities between   \pp and
heavy-ion interactions. Such comparison requires data with high
statistics on many particle species and particularly on
multi-strange baryons. We are convinced that the LHC will provide a
very deep insight into strangeness production dynamics  in \pp and
in heavy-ion collisions and it will allow us to verify or strengthen 
our conclusions drawn from the existing data.

\vskip 0.5cm

We acknowledge the support from the German BMBF. K.R. acknowledges
supports of the Polish Ministry of Science MEN and the Alexander
von Humboldt Foundation. The financial support of the DFG-NRF and
the South Africa--Poland scientific collaborations are also
gratefully acknowledged.

\end{document}